\documentclass[reprint,twocolumn]{revtex4-1}

\usepackage{graphicx}
\usepackage[sort&compress]{natbib}
\usepackage{vaucanson-g}
\usepackage{amssymb}

\begin{document}

\title{Effective Theories for Circuits and Automata}
\author{Simon DeDeo}
\email{{\tt simon@santafe.edu}}
\affiliation{Santa Fe Institute, Santa Fe, NM 87501, USA}
\date{\today}

\begin{abstract}
Abstracting an effective theory from a complicated process is central to the study of complexity. Even when the underlying mechanisms are understood, or at least measurable, the presence of dissipation and irreversibility in biological, computational and social systems makes the problem harder. Here we demonstrate the construction of effective theories in the presence of both irreversibility and noise, in a dynamical model with underlying feedback. We use the Krohn-Rhodes theorem to show how the composition of underlying mechanisms can lead to innovations in the emergent effective theory. We show how dissipation and irreversibility fundamentally limit the lifetimes of these emergent structures, even though, on short timescales, the group properties may be enriched compared to their noiseless counterparts. [Appearing in Chaos Special Issue on Randomness, Structure and Causality, ed. Jim Crutchfield and Jon Machta. \emph{Chaos} 21, 037106 (2011)]
\end{abstract}

\maketitle

\noindent {\bf Many systems, especially those with large numbers of underlying degrees of freedom, are best described by what are known as effective theories. These theories allow us to describe the relevant high-level phenomena while remaining ignorant, or at least agnostic, about the fine-grained details of a system's state. Here, we show how to construct effective theories of phenomena that may show irreversibility or dissipation, a question particularly relevant for biological, social and computational systems. We use a hierarchical decomposition technique from semigroup theory that allows one to take finite state automata (standard, if restricted, models of computation) and determine the irreducible components of the process. Truncating the hierarchy provides an effective theory. We study how different underlying mechanisms lead to qualitatively different effective theories, and show how noise and irreversibility interact to produce new computational phenomena on short timescales, which, in the presence of irreversibility, eventually dissipate at longer intervals.}

\section{Introduction}

Natural systems usually have at least two layers of description. At the lower level is that of \emph{mechanism}, a description of the relevant physical (chemical, biochemical, \&c.) properties and how they change, deterministically, over time: either by laws relating properties, or by laws relating probability distributions over properties. This lower level need not be completely reduced -- for example, the behavioral regularities of a representative person, calibrated by laboratory experiments, might provide an adequate lower-level theory. 

At a level ``higher than'', or above, this description, is often an \emph{effective theory} of the phenomena at hand. The individuals in a crowd follow basic psychological laws, say -- but the crowd itself shows new properties (perhaps density waves, stream-crossings, or a Maxwellian velocity distribution.) One sign of a complex system is when the higher-level description is not a simple transformation of the lower-level description -- when the effective theory at the higher level is not simply (say) identical to that of the lower level dynamics with shifted parameters. Crowds and swarms may be more, or less, intelligent than the individuals that compose them, and may have incommensurate abilities.

The intention here is to demonstrate this phenomenon for effective theories of aggregations of Boolean circuits. The circuits that form the mechanism level have long been models for information processing in natural systems~\cite{vN56,Kauffman:1973vz}, and have been chosen for their familiarity. Their directed, acyclic graph structure makes them a natural way to describe the basic units of a computational process unfolding in time. When noisy operation and feedback is allowed, they show dissipative properties, and can be described as finite-state probabilistic automata. It is in this latter language that we will phrase the higher-level effective theory.

Our choice of probabilistic automata allows the use of certain mathematical tools, unavailable in more general cases such as computation by a Turing machine, to quantify the characteristic features of the effective theory. In particular, within our construction, ``more is different''~\cite{more} can be quantified as the emergence of new, irreducible group structures.

We first consider the non-probabilistic case. We briefly describe how circuits with feedback are qualitatively different from the acyclic case, and then introduce the finite state automata. We then show how automata can be wired together, or composed, in an operation analogous to the interconnection of gates. We  introduce a crucial theorem, the Krohn-Rhodes decomposition, and show how it relates to the underlying structures of the Boolean circuits the automata describe.

This will give us the necessary framework to show the emergence of effective theories in non-probabilistic systems. In particular, it will allow us not only to decompose theories so that they might be compared to each other, but also to show how they can be coarse-grained~\cite{rhodes10}. It is this coarse-graining process that allows us to define effective theories and to fairly compare the properties of microscopic and macroscopic dynamics; it provides a smoothed, lower-dimensional, computational process.

We then show how these results can be extended to the probabilistic domain. We show how to separate the random and non-random components of such a system. This provides a solvable decomposition of the most general probabilistic computation -- the ``Bernoulli Turing Machine''~\cite{jim}. Formally, the introduction of probabilistic behavior leads to a non-random component with more complicated group structures and the possibility of both irreversibility and dissipation. Transient behavior may be associated with larger group structures than the corresponding noiseless machine. However, this window finally closes, and the long-timescale effect of noise is to destroy all these structures, including those of the noiseless counterpart.

\section{From Circuit to Automaton}

Logical circuits -- compositions of basic operations (gates) such as ${\tt AND}$, ${\tt OR}$ and ${\tt NOT}$ -- are a natural way to represent information processing and computation in natural systems~\cite{Bonn:2008p18667,Davidson:2002p18651,Davidson:2011p18642,Macia:2009p18657}.

For the case of Boolean functions without feedback, their algebraic properties are particularly clean, and provide a handle on how new functional classes emerge on composition. For example, ${\tt AND}$ and ${\tt OR}$ together can only represent a fraction of all possible functions, while ${\tt NAND}$ or ${\tt NOR}$ alone can cover the entire space. The full description of these relations is the Post lattice~\cite{post20,lau06}, which shows how functional classes implemented by different combinations of fundamental gates (logical bases) enclose or are enclosed by others.

Because their underlying structures are directed, acyclic graphs, they can also be seen as descriptions of causal processes, in the manner of Pearl's causal networks~\cite{pearl}. The classes that make up the Post lattice then amount to different ways of partitioning the space of causal processes -- once any loops are ``unrolled'' so as to eliminate cycles.

\begin{figure}
\includegraphics[width=1.5in]{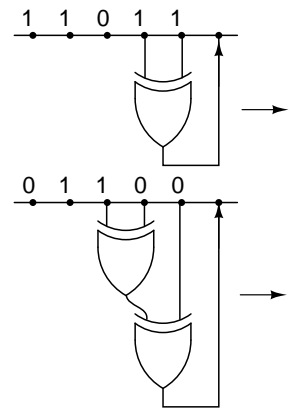}
\caption{Boolean functions on the two and three-bit register. At each time step, the device moves one step to the right, inducing feedback. These machines can be identified with logical nets; the acyclic graphs that compose them are the fundamental mechanism in our demonstration.}
\label{boolean}
\end{figure}

In many cases, however, one wants a simultaneous account of causality \emph{and} feedback (or self-action, joint regulation, and so forth.) The extension of Boolean circuits to this case is natural -- one simply takes outputs from later in the process, and connects them to inputs upstream. This is often described using the formalism of logical nets (see, \emph{e.g.}, Ref.~\cite{dassow05}); Fig.~\ref{boolean} shows two examples of the particular cases we will examine; the different ways a limited number of logical gates can be combined are the structures we associate with the mechanism layer.

The seemingly minor change of allowing feedback leads to -- among other things -- an uncountably infinite number of ``precomplete'' classes (\emph{i.e.}, classes just short of full coverage of the space of all feedback functions)~\cite{dassow81,dassow05}. The breakdown of the classification by logical nets leads us to formulate a higher-level theory in terms of finite state automata~\cite{pippenger97}. We introduce these machines, and some of their mathematical properties central to our discussion, in the next section.

\section{Automata and Semigroups}

\begin{figure}
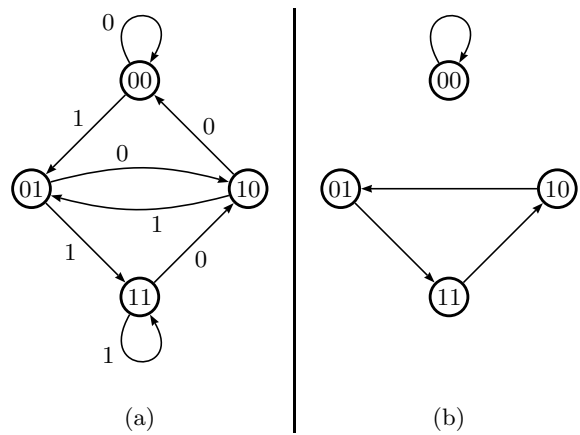

\begin{tabular}{c|c}

\VCDraw[0.8]{
\begin{VCPicture}{(1,0)(9,11)} 
\State[00]{(5,9)}{00}

\State[01]{(2,6)}{01}
\State[10]{(8,6)}{10}

\State[11]{(5,3)}{11}

\EdgeR{00}{01}{1}
\EdgeR{01}{11}{1}
\EdgeR{11}{10}{0}
\EdgeR{10}{00}{0}

\ArcL{01}{10}{0}
\ArcL{10}{01}{1}
\LoopN{00}{0}
\LoopS{11}{1}
\end{VCPicture}
} &

\VCDraw[0.8]{
\begin{VCPicture}{(1,0)(9,11)} 
\State[00]{(5,9)}{00}

\State[01]{(2,6)}{01}
\State[10]{(8,6)}{10}

\State[11]{(5,3)}{11}

\EdgeR{01}{11}{}
\EdgeR{11}{10}{}
\EdgeR{10}{01}{}

\LoopN{00}{}
\end{VCPicture}
} \\
(a) & (b)
\end{tabular} 
\caption{Left: the two-bit left-shift automaton. Right: the {\tt XOR} function, $t_i \rightarrow t_{i-2} {\tt ~XOR~} t_{i-1}$, corresponding to the top mechanism of Fig.~\ref{boolean}.}
\label{monoid}
\end{figure}

We consider first the non-probabilistic case. The simple deterministic~\footnote{We use the word `deterministic' in the automata sense -- \emph{i.e.}, a deterministic system is such that a letter-state pair is associated with a unique transition, $\delta:\Sigma\times Q\rightarrow Q$ as opposed to $\delta:\Sigma\times Q\rightarrow \mathrm{Powerset}[Q]$. We do not consider non-deterministic automata explicitly in this paper; well-known methods allow one to construct a deterministic automata corresponding to any non-deterministic one. We use the word `probabilistic' to refer to situations in which a deterministic transition is chosen from a set  probabilistically.} automaton we define by the tuple $(\Sigma, Q, \delta, q_0)$, where $\Sigma$ is the input alphabet, $Q$ the set of internal states and $\delta:\Sigma\times Q\rightarrow Q$, a map from the current state and the current input letter to a subsequent state -- $\delta(s,q)=q^\prime$ means that when the automaton is in state $q$ and receives the input letter $s$ (in $\Sigma$) it will transition to state $q^\prime$.

The state $q_0$ is the initial state of the system; since we will be concerned with processes that run forever, and not with the language-recognition properties of automata, we will not define terminal states.

An important case for the description of natural systems is when the input alphabet has only one element. This single-letter environment amounts to a ``passage of time'' operator, advancing the system from one state to the next. In other cases, different letters in the input alphabet will amount to different environmental signals, altering the transitions the system makes between states.

The other important structure for our analysis is the semigroup. A semigroup is a set $S$ of elements that can be ``multiplied'' together to produce other elements in the set. Multiplication is associative and $a\cdot(b\cdot c)$ gives the same result as $(a\cdot b)\cdot c$. In some cases, a semigroup will have an identity element, $\iota$ -- for any element $s$ in the semigroup, $s\cdot\iota=s=\iota\cdot s$. A semigroup with an identity element is called a monoid. The differences between semigroups and monoids will not be essential in our discussion.

In the case where there are inverses -- \emph{i.e.}, when for every element of the set, $s$, there is another element, $s^\prime$, such that $s\cdot s^\prime$ is the identity $\iota$ -- the semigroup is a group. In the physical sciences especially, groups have been used to describe symmetries of a system -- for example, the group of rotations, or reflections, or groups such as hypercharge associated with the insensitivity of physical processes to internal properties of a particle. Semigroups, we shall see, are associated with a wider range of processes, including irreversible processes~\footnote{As with `non-deterministic' and `probabilistic,' it is worth distinguishing `irreversible' and `dissipative'. Irreversibility need not be associated with any randomness, and can be a consequence of a completely non-probabilistic map, such as an irreversible computation that deletes input. Dissipation, by contrast, we associate with processes that can not be run backwards because different histories have been merged due to an ignorance of which randomly-chosen operation was performed. We return to this distinction at the end of Sec.~\ref{correlated_noise}.} where there is no inverse operation that takes everything back to a prior state.

The \emph{free semigroup} on a letter, or set of letters, is defined as the set of all compositions of the letters -- for example, the free semigroup on $\Sigma=\{a\}$ is the (infinitely large) set $\{a,aa,aaa,aaaaa,\ldots\}$; this is often denoted $\Sigma^+$. By identifying the results of different compositions, one produces smaller, often finite, semigroups. For example, by setting $aaaa$ equal to $a$, one gets the monoid $M$ consisting of $\{a,aa,aaa\}$, where $aaa$ now appears as the identity element. Note that in this case, $M$ is also a group -- the inverse of $a$ is $aa$, the inverse of $aa$ is $a$, and $aaa$ is its own inverse -- and is called the cyclic group of order three; it counts inputs modulo three, and is written $Z_3$.

Another important semigroup is the \emph{full transition semigroup} on $n$ states. It is the largest semigroup on $n$ states, and has $n^n$ elements, each of which amounts to a map from the set of states to itself. Included in the full transition semigroup are all of the permutations of the set, as well as the ``irreversible'' operations that may map many inputs to the same element.

The correspondence between finite state automata and semigroups (\emph{i.e.}, groups with neither inverses nor an identity element) allows one to draw on the tools of abstract algebra. Consider the automaton $(\Sigma, Q, \delta, q_0)$. A string of input letters is a member of $\Sigma^+$. The transition function $\delta$ can be extended to accept elements of $\Sigma^+$ in the natural manner -- by definition, any $s\in\Sigma^+$ is a string $s_1s_2\cdots s_n$, with each $s_i$ in $\Sigma$. Then, $\delta(s,q)$ is $\delta(s_n,\ldots\delta(s_{2},\delta(s_1,q))\ldots)$ -- \emph{i.e.}, the state you end up in after receiving $s_1$ in state $q$, going to the new state $\delta(s_1,q)$, receiving letter $s_2$, and so forth.

Given this extension of $\delta$, we define an equivalence relation, $\equiv$, on members of $\Sigma^+$ by
\begin{equation}
a\equiv b \textrm{~if and only if~} \delta(a,q)=\delta(b,q)~\forall q\in Q.
\end{equation}
In other words, two members of $\Sigma^+$ are equivalent if and only if they have the same effect on the automaton regardless of the state the automaton begins in. This structure is called the transition semigroup, and makes the correspondence between automata and semigroups complete.

Semigroups often have subsets that turn out to be other semigroups, or even groups. Note that for a subset $S^\prime$ of a semigroup $S$ to be a semigroup, it has to be closed under multiplication -- in other words, multiplying together any two elements of $S^\prime$ has to give you another element in $S^\prime$. The monoid $M$, above, has no non-trivial subgroups (or subsemigroups.)

Fig.~\ref{monoid}a shows a more interesting example: the ``left-shift'' automaton that has an input alphabet $\Sigma=\{0,1\}$, and four states. This machine describes a system with a two-bit internal register, that processes new elements from the environment by shifting the register bits the left and appending the input bit. Such a machine does not have a group structure: whatever state the system is in at time $t$, after accepting two more inputs, it has completely forgotten its prior state. There is no way to undo this operation, and all past information is erased. The automata thus induces a semigroup with four single-element subsemigroups, and no non-trivial subgroups~\footnote{For clarity, we do not include the transitory $0$ and $1$ states in our automata pictures.}.

Fig.~\ref{monoid}a forms a base from which one can define other functions. Consider, for example, the process $t_{i}\rightarrow t_{i-2} {\tt~XOR~} t_{i-1}$. This can be described by a finite state automaton whose skeleton is the left-shift semigroup; in particular, the input set $\Sigma$ is now the one letter passage of time operator, and the transition structure, Fig.~\ref{monoid}b, chooses for each state one of the two outgoing arrows of Fig.~\ref{monoid}a. Note that while the left-shift register has no non-trivial subgroups, the {\tt XOR} machine does -- in particular, syntactic monoid is $Z_3$.

There are $2^{(2^n)}$ single-letter automata that can be defined on the n-bit left-shift register in this fashion, one for each of the possible Boolean truth tables $\mathcal{B}^n\rightarrow\mathcal{B}$. We can merge these machines together to produce the \emph{left-shift powerset} automaton with $2^{(2^n)}$ input letters, each letter referring to a transition implemented by one of the single-letter automata.

The left-shift powerset automaton is much smaller than the full transformation semigroup, which has $(2^n)^{(2^n)}$ elements. However, when transitions on the automata are composed -- \emph{i.e.}, when one compares the semigroup of the left-shift powerset with the full transition semigroup -- then the two are isomorphic. We prove this in the Appendix.

\section{Automata Cascades}

At the mechanism level in our demonstration, composition is simply the wiring together of logical gates in structures of increasing complexity, as in Fig.~\ref{boolean}. Such compositions are relatively simple to understand -- one connects outputs from one circuit to the inputs of a another, producing a directed acyclic graph.

In many cases in both natural and engineered systems, we would like our environmental response -- the transitions our machine makes -- to be modulated by other processes, including other machines. Indeed, the ability for one machine to control another seems central not only to the notion of computation~\footnote{A related notion is that of one machine ``pre-processing'' the environment and handing off its output to the next machine -- this turns out to be equivalent to the description we use here, but harder to formulate in terms of our finite state automata.}, but to collective phenomena in general.

A \emph{cascade} of automata is a restricted form of such control. It is a feed-forward system, with a machine at level $n$ receiving a set of letters from the $n-1$ machines above it, and the environment itself. The machine at the top level (level 1) receives only the environmental input $s$, and makes a transition from state $q_1$ to $q_1^\prime$; the machine at level 2 receives a joint symbol, $(s,q_1)$ and makes a transition from $q_2$ to $q_2^\prime$; the machine at level 3 receives the triple $(s,q_1,q_2)$ and so forth.

\begin{figure}
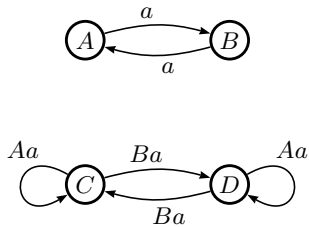

\VCDraw[0.8]{
\begin{VCPicture}{(0,0)(8,8)} 
\State[A]{(2,6)}{A}
\State[B]{(6,6)}{B}

\State[C]{(2,2)}{C}
\State[D]{(6,2)}{D}

\ArcL{A}{B}{a}
\ArcL{B}{A}{a}

\ArcL{C}{D}{Ba}
\ArcL{D}{C}{Ba}
\LoopW{C}{Aa}
\LoopE{D}{Aa}

\end{VCPicture}
}
\caption{A cascade of two $Z_2$ counters; the lower-level machine has two possible transitions, depending on the state of the higher-level machine. The cascade now can count modulo four, with the high-level machine counting even versus odd, and the lower-level machine tracking the second bit.}
\label{cascade}
\end{figure}

A very simple example of a cascade is show in Fig.~\ref{cascade}. Here, two machines, both equivalent to a $Z_2$ counter, are cascaded. The top-level machine accepts a single input letter, $a$, from the environment; the lower-level machine receives a joint signal $Aa$ or $Ba$, depending on the state of the machine above. Inspection shows that this cascade of two modulo-two counters allows us to count modulo four.

Cascades of automata are essentially hierarchical. Machines higher up the chain control more general features of the system. They require fewer inputs, and are ignorant of the finer-scale details of state structures below. Their states are effectively superstates -- collections of states of the equivalent collapsed machine. The top-level states $A$ and $B$ of Fig.~\ref{cascade}, for example refer to pairs of states in the equivalent $Z_4$ counter.

It is in these senses that cascades provide a graded set of effective theories. Truncating a cascade at some level allows one to ``smooth'' a complicated, multi-state process in a manner consistent with its internal logic.

Cascades have a particularly special form. In the next section, however, we will see that \emph{any} automaton can be rewritten in this way, with each level composed of particularly simple subroutines.

\section{The Krohn-Rhodes Decomposition}

In some cases, as we saw in the previous sections, the semigroup associated with an automaton turns out to be a group. The Cayley theorem -- that all groups are isomorphic to subgroups of the permutations -- makes these automata simple to identify. In particular, for such an automaton every transformation induced by a letter of $\Sigma$ leads to a permutation of the states. 

At the other extreme are the ``resets.'' A letter $s$ in $\Sigma$ is a reset if it has the same effect on all states -- if it resets the machine, in other words, to a unique state.

In between, of course, are operations that permute some states and reset others -- for example, a letter $s$ acting on a subset of states might act as a permutation, but on a different set it acts as a reset, taking all elements to a single final state. Examples of these mixed machines include that of Fig.~\ref{decomp}b. Conversely, ``pure'' machines -- those that are combinations of only permutations and resets -- are called permutation-reset machines. A particularly simple example of a permutation-reset machine is the flip-flop -- a two-state machine with two resets and the identity operation.

\begin{figure}
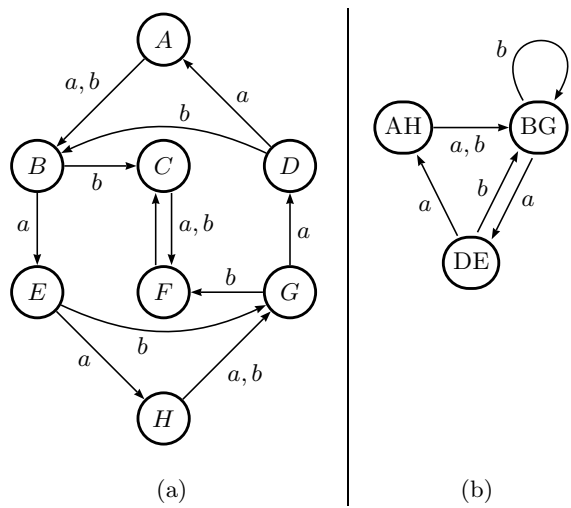

\begin{tabular}{c|c}
\VCDraw[0.7]{
\begin{VCPicture}{(-1,-1)(9,13)} 

\LargeState\State[A]{(4,12)}{000}

\State[B]{(0,8)}{001}
\State[C]{(4,8)}{010}
\State[D]{(8,8)}{100}

\State[E]{(0,4)}{011}
\State[F]{(4,4)}{101}
\State[G]{(8,4)}{110}

\State[H]{(4,0)}{111}

\EdgeR{000}{001}{a,b}
\EdgeR{001}{010}{b}
\VArcR{arcangle=-25}{100}{001}{b}
\EdgeR{100}{000}{a}

\ForthBackOffset \EdgeL{010}{101}{a,b} \EdgeL{101}{010}{} \RstEdgeOffset

\EdgeR{001}{011}{a}
\EdgeR{110}{100}{a}

\VArcR{arcangle=-25}{011}{110}{b}
\EdgeR{110}{101}{b}

\EdgeR{011}{111}{a}
\EdgeR{111}{110}{a,b}

\end{VCPicture}
} ~~  & \VCDraw[1.0]{
\begin{VCPicture}{(0.5,-1.5)(5.75,5)} 

\LargeState

\StateVar[\mathrm{~AH~}]{(1.5,4)}{AH}
\StateVar[\mathrm{~BG~}]{(4.5,4)}{BCG}
\StateVar[\mathrm{~DE~}]{(3,1)}{EDF}

\EdgeR{AH}{BCG}{a,b}
\EdgeL{EDF}{AH}{a}

\ForthBackOffset
\EdgeL{BCG}{EDF}{a}
\EdgeL{EDF}{BCG}{b}
\RstEdgeOffset

\LoopN{BCG}{b}

\end{VCPicture}
}  \\ \\
(a) & (b) \\
\end{tabular}
\caption{Effective theories and ``smoothed'' computations. (a) An eight-state automaton with two input letters, built from ${\tt XOR}$ gates on the left-shift register. (b) The top level of the decomposed machine. Despite the complicated internal structure of (a), at the coarse-grained level, the process is seen count modulo three (on receiving $a$ signals), and reset (on receiving $b$ signals.) Movement within the three superstates of (b) is dictated by $Z_2$ counters and resets at lower levels in the decomposition. For example, states C and F are shared by all three states of the high level theory; transitions to this sub-space are described by resets at lower levels.}
\label{decomposition}
\end{figure}

The surprising result of the Krohn-Rhodes theorem~\cite{kr,rhodes10} is that \emph{any} finite state automaton can be decomposed into a cascade whose automata contain only the simple groups and the flip-flop. This cascade produces behavior that maps homomorphically onto to the original machine~\footnote{Homomorphic behavior means that there exists a map between the two semigroups that preserves multiplication. There is a homomorphic mapping $\phi$ between two groups in the case that multiplication is preserved: $\phi(a)\cdot\phi(b)=\phi(a\cdot b)$. Homomorphic mappings need not one-to-one, and in general the cascade will be larger than the original machine; in particular, the mapping may need to be restricted to a subautomaton of the cascade, and some states within this subautomaton may need to be identified with each other.}. In contrast to Turing-complete processes, one can not only can define a ``restricted'' algorithmic information complexity for the finite state automata by standard minimization techniques, but can also break them down into a small number of hierarchically organized subroutines.

Ref.~\cite{kr} proved the existence of such decompositions, but did not present a simple way to find them. The holonomy decomposition~\cite{eilenberg76,holcombe04,Nehaniv:2000p19116,Maler:2010p19157} provides an efficient means of reconstruction; computations in this paper use an implementation of the holonomy decomposition in the ${\tt sgpdec}$ package~\cite{sgpdec,EgriNagy:2008p19414,attila09} in ${\tt GAP}$~\cite{GAP4}.

An example of the decomposition is shown in Fig.~\ref{decomposition}. The complicated structure of Fig.~\ref{decomposition}a breaks down into a three-level cascade. The top level, shown in Fig.~\ref{decomposition}b is a $Z_3$ counter with a reset; the lower two levels include additional $Z_2$ counters. The ``superstates'' of Fig.~\ref{decomposition}b are not disjoint: all three of the superstates include states C and F.

One of the interesting features of the Krohn-Rhodes theorem is the difference in treatment of the reversible (group symmetry) and irreversible transformations of the system. While the groups resolve themselves into a non-trivial catalog of simple subunits, there are no ``irreducible'' semigroups of dissipation beyond the flip-flop. There are distinct and irreducible groups of reversible computations, but the irreversible aspects of a computation decompose finally into collections of pure identity-resets.

We now have all of the necessary tools to demonstrate the emergence of incommensurate effective theories in non-probabilistic systems.

\section{Effective Theories in Non-Probabilistic Systems}

Fig.~\ref{monoid}b shows the action of an ${\tt XOR}$ structure on two bits. It contains a $Z_3$ counter (as in Fig.~\ref{decomposition}, the superstates of the top level overlap.) Such a counter can be chained in different ways; with an appropriate choice of cascade, it can play the role of a multi-digit trinary counter (and similar, degenerate machines.)

The apparatus of group theory, however, provides a clear answer to what the machine can \emph{not} do and constrains the form of \emph{any} effective theory built out of parallel or serial compositions of these machines. The group structures of these new effective theories must be (Jordan-H\"{o}lder) reducible to their constituents.

By contrast, combinations of gates -- as opposed to their induced automata -- will not be so restricted, and it is these innovations, at the mechanism level, that we will be concerned with. In general, combinations of larger numbers of circuits, with greater environmental sensitivity and over longer histories will naturally produce more complicated behavior. 

This is similar to how a collection of $n$ members of a crowd will be more complicated than a single individual, because one now has to keep track of $n$ times as many variables, as well as the combinatorics of their interactions. However, depending on the nature of the interactions, the effective theory of such a collection -- a theory over a set of states comparable in size to that of the original members -- may have qualitatively new properties. A pair of $Z_2$ counters may be cascaded to count modulo 4, but the mechanisms underlying those counters may be re-wired in such a way to produce a machine with novel decomposition.

The only permutation machines with a single input letter -- corresponding to groups with a single generator -- are the counters, and they factor into the cyclic groups of prime order (if a machine can count modulo three, copies can be chained together in a cascade to count modulo 27 ($3^3$), but can not count modulo 7, for example.) We will see that mechanism-level compositions, by contrast, lead to new categories of machine that can count modulo an arbitrary number (and whose cascades can do arithmetic in an arbitrary combination of bases.)

New and irreducible effective theories may also arise if one allows the machine to discriminate different environmental states -- \emph{i.e.}, if one expands the input alphabet. Since many groups have a small number of generators (and the simple groups have at most two) one does not need a great amount of environmental diversity to realize these more complicated structures.

\begin{table}
\begin{tabular}{c}
\begin{tabular}{l|c|c}
3-bit {\tt XOR} & 1 letter & 2 letters \\ \hline
1 gate & $Z_{n=7}$ & PSL$(3,2)$ \\
 2 gates & $Z_{n=2,3}$ & $S_4$ \\
 3 gates & -- & -- \\
 \end{tabular}
\\ \\
\begin{tabular}{l|c|c}
4-bit {\tt XOR} & 1 letter & 2 letters \\ \hline
1 gate & $Z_{n=2,3,5}$ & $A_8$ \\
 2 gates & $Z_{n=7}$ & PSL$(3,2)$ \\
 3 gates & -- & $S_{n=3,5,6}$  \\
 4 gates & -- & -- 
\end{tabular}
\end{tabular}
\caption{The relationship between mechanism, environment, and top-level emergent group structures, with increasing complexity in mechanism (going down the table) and environmental diversity (from a singleton environment to a binary signal.) Notation: $Z_n$ (counters modulo $n$); $A_n$ (the alternating groups on $n$ elements); $S_n$ (full permutation group); PSL$(3,2)$ (second smallest non-Abelian simple group after $A_5$.) In many cases, these groups appear as semidirect products with other groups (not shown on table, for clarity.) A dash indicates that no new simple groups appear.}
\label{xorcharts}
\end{table}

We show a specific example of these effects in Table~\ref{xorcharts}, where we consider ${\tt XOR}$-based mechanisms on three- and four-bit registers. In particular, each entry for the ``1-letter'' column in the table shows the new top-level groups that are possible by use of at least one of the Boolean circuits with the specified number of inputs and ${\tt XOR}$ gates, operating in the left-shift register fashion exemplified by Fig.~\ref{boolean}. We consider circuits with gates of fan-in two and unlimited fan-out, which give a total of 610 distinct truth tables found by enumeration of the 4643 topologically distinct graphs for the mechanism complexities listed in the table.

The standard left-shift register has a singleton environment -- the passage of time operator. In the case of the 2-letter environment, one enlarges the systems under consideration to pairs of circuits, with two possible input letters that choose which of the circuit structures to implement at each step.

The composition of increasing numbers of gates leads to a number of new and irreducible group structures. Underlying mechanisms of limited universality ({\tt XOR}, for example, is not even a complete basis in the Post lattice) lead not only to new effective theories, but to a profligate number of them, including many near-maximal symmetries on the full state set. We show in our table only the simple groups, but in many cases these combine in more complicated structures:  for example, $(Z_2\times Z_2\times Z_2)\rtimes \mathrm{PSL}{(3,2)}$ can be implemented with a pair of circuits, with two gates each, on the 4-bit register.

Since (as shown in the Appendix) the left-shift powerset is isomorphic to the full transformation semigroup, when all truth tables become accessible (when the underlying gates are drawn from a complete basis), then \emph{all} groups become accessible, given sufficient environmental diversity and mechanism complexity. 

Complete coverage of the full $\mathcal{B}^n\rightarrow\mathcal{B}$ space is not necessary to attain the full transformation semigroup. For example, even though one requires circuits of at least nine ${\tt NAND}$ gates to cover all 3-bit truth tables, one can achieve the full transformation semigroup on eight states with mechanisms that need at most only five ${\tt NAND}$ gates.

Despite this, such compositions at the mechanism level will still become increasingly vulnerable to noise in the individual elements of the system. When gates become unreliable, undesired transitions may occur, and compositions of gates exacerbate this problem by providing multiple failure-points. As we shall see in the next two sections, the presence of even a small amount of noise leads to an irreversible degradation of the group structures of the effective theory.

\section{Probabilistic Automata}

The Krohn-Rhodes decomposition applies to deterministic, non-probabilistic systems, and is not immediately applicable to situations in which the effect of an input symbol is probabilistic. In this section, we show how to break up such probabilistic machines into a cascade, of a probabilistic process at the top (a ``Bernoulli automaton''), which feeds a symbol, chosen from a distribution, to a non-probabilistic automaton beneath. This construction was first demonstrated by Ref.~\cite{Maler:1995p19164}.

An $n$ state probabilistic automaton is defined by $(\Sigma, Q, p, q_0)$, where $p:Q\times\Sigma\times Q\rightarrow[0,1]$,  replaces the state transition function. $p(q_1,s,q_2)$ gives the probability that when the machine is in state $q_1$, and receives the input letter $s$, it makes a transition to $q_2$. The probability is normalized so that, for an $n$ state automaton, $\sum_{i=1}^n p(q,s,q_i)$ is unity for all $s\in\Sigma$, $q\in Q$.

Consider the space of all possible transitions on $n$ states -- the full transformation semigroup, $M$. There are $n^n$ such transitions in the semigroup, which can be written as functions $m(q)$. A probabilistic automaton associates a probability, $\pi$, with every $m$, and input letter, $s$, as follows:
\begin{equation}
\pi{(m, s)}=\prod_{i=1}^n p[q_i, s, m(q_i)].
\label{prob}
\end{equation}
In words, the probability of $m$ given $s$ is the product of the probabilities that each arrow in the original automaton corresponding to $a$ is obeyed.

The decomposition then proceeds as expected. A Bernoulli automaton receives signal $s$ from the environment, and chooses $m\in M$ according to the distribution of Eq.~\ref{prob}. A deterministic machine with the same $n$ states as the original probabilistic machine, but which now accepts up to $n^n$ letters, then executes the transition $m$. This latter machine can be decomposed in the standard way. 

\begin{figure}
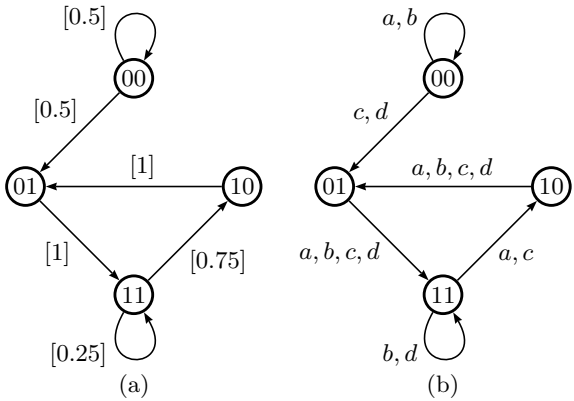

\begin{tabular}{cc}

\VCDraw[0.8]{
\begin{VCPicture}{(1,1)(9,11)} 
\State[00]{(5,9)}{00}

\State[01]{(2,6)}{01}
\State[10]{(8,6)}{10}

\State[11]{(5,3)}{11}

\EdgeR{00}{01}{[0.5]}
\EdgeR{01}{11}{[1]}
\EdgeR{11}{10}{[0.75]}
\EdgeR{10}{01}{[1]}

\LoopN{00}{[0.5]}
\LoopS{11}{[0.25]}

\end{VCPicture}
} & \VCDraw[0.8]{
\begin{VCPicture}{(1,1)(9,11)} 
\State[00]{(5,9)}{00}

\State[01]{(2,6)}{01}
\State[10]{(8,6)}{10}

\State[11]{(5,3)}{11}

\EdgeR{00}{01}{c,d}
\EdgeR{01}{11}{a,b,c,d}
\EdgeR{11}{10}{a,c}
\EdgeR{10}{01}{a,b,c,d}

\LoopN{00}{a,b}
\LoopS{11}{b,d}

\end{VCPicture}
} \\
(a) & (b)
\end{tabular} 
\caption{Decomposition of a (single input letter) probabilistic automaton into a 4-letter deterministic machine; $\pi(a)=\pi(c)=3/8$; $\pi(b)=\pi(d)=1/8$.}
\label{decomp}
\end{figure}

As an example, consider introducing noise to the automaton of Fig.~\ref{monoid}b to produce the probabilistic automaton defined by the truth table $[0.5, 1, 1, 0.75]$ -- where the numbers now refer to the probability of emitting a one. The probabilistic version is shown in Fig.~\ref{decomp}a, and the structured, non-random part in Fig.~\ref{decomp}b.

Eq.~\ref{prob} is a very general -- but not the only -- noise model for the kinds of computational process  described above; it corresponds to the $\epsilon$-independent failure model~\cite{vN56,Pippenger:1989p18957}. Other models are possible, depending on one's beliefs about how the computation departs from non-probabilistic behavior; we return to this question in Sec.~\ref{correlated_noise}.

\section{Emergence in Probabilistic Systems}
\label{emergence_prob}

The construction of the previous section shows that adding probabilistic behavior to a finite state automaton generally increases the complexity of the underlying deterministic mechanism. When introducing noise to the machine of Fig.~\ref{monoid}b to produce Fig.~\ref{decomp}a, for example, the underlying structure goes from a singleton alphabet to a four-letter alphabet, and the holonomy decomposition goes from a single $Z_3$ counter to a four-level cascade of $Z_3$, $Z_2$ and identity-reset machines.

In the extreme case, a uniformly-distributed probabilistic choice over the full transition semigroup is the most complex machine -- in terms of the size of the irreducible groups. And yet when coupled to a uniform Bernoulli process, it produces a purely random stream of numbers. The larger the structure one moves around in, the closer one gets to randomness; or, put another way, the more features of the environment a machine is sensitive to, the more random that machine appears when the relevant variables are unobserved.

The generators of the left-shift powerset will in general have different probabilities. While it is tempting to describe the resulting machine as probabilistic draw of the different semigroups (and their Krohn-Rhodes decompositions) associated with each single-letter automaton, this elides a crucial distinction between the automaton and its expansion to the semigroup structure. For example, the semigroup associated with the ${\tt XOR}$ automaton  includes the cyclic group $Z_3$; however, the ``shift by one'' and ``shift by two'' elements of that group will have different probabilities when noise is introduced, since they require different compositions of generators.

In particular, as the machine unfolds in time, different elements of the semigroup will appear with different probabilities. While the generators are themselves members, other elements of the full transition semigroup require multiple compositions, and so the probability distribution over the different elements will be a function of time. 

The asymptotic distribution depends on the presence of irreversible transformations. When resets are present, one finds that the limiting distribution is concentrated solely at these points. Under the failure model of Eq.~\ref{prob}, noisy machines that run for ever forget everything.

This is most simply visualized as the limit of a random walk on the directed graph of the semigroup, where each vertex corresponds to a member of the semigroup and each directed edge corresponds to some multiplication by an element of the left-shift powerset -- this graph is analogous to the Cayley graph for groups. Convergence depends on the details of the noise mechanism, but the resets form a strongly connected subgraph that, once reached, is never left.

\begin{figure}
\includegraphics[width=3.25in]{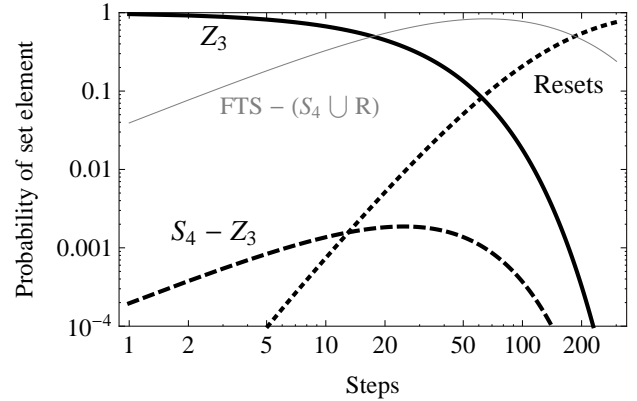}
\caption{The rise, and fall, of group structures in noisy Boolean systems. For the ${\tt XOR}$ function of Fig.~\ref{monoid}b, with a 1\% bit-flip rate, we show the probability that the function implements one of the group elements of the noise free case ($Z_3$, solid line), one of the four reset automata (dotted line), or any group element not in $Z_3$ (the set $S_4-Z_3$, dashed line.) The light gray line shows the remainder. As time passes and errors accumulate, the chances of a transformation being a member of $Z_3$ decline, while the reset automata come to dominate the long term input-output maps. The non-trivial relationship between noise and complexity can be seen when, for a brief period, representatives of more complex groups appear and even dominate the pure resets.}
\label{decay}
\end{figure}

Noisy automata, then, can show complexity only on short timescales. This is shown in Fig.~\ref{decay}, where the shifting distribution of semigroup structures is shown for the noisy-${\tt XOR}$ (\emph{i.e.}, the truth table $[\epsilon,(1-\epsilon),(1-\epsilon),\epsilon]$.) As noise increases, the machine becomes less likely to realize any element of the $Z_3$ cyclic group. The ultimate limit of the process is to end up at the resets. This amounts to a forgetting, and a time horizon of complexity.

For a brief window, members of other groups (elements of the set of all permutations, minus the members of $Z_3$) appear. These other members are expected, since they are included in the left-shift semigroup; however, their appearance is only fleeting, and (in this case) never becomes a significant driver of dynamics. The question of whether such noise-assisted group structures can become significant is open, and may provide insight into the use and management of noise in biological systems (see, \emph{e.g.}, Ref.~\cite{Krakauer:2002p18958,Rao:2002wz,Suel:2007vz,Lehner:2008wu, BarEven:2006uu, Hornung:2008tf}; we return to this question in Sec.~\ref{correlated_noise}.)

One can now see the two-fold relationship between noise and the increasing numbers of group and semigroup operations.

Combining underlying mechanisms in the non-probabilistic case leads to a growing diversity of group structures; with sufficiently complex environments, and sufficiently large mechanisms, one can instantiate every classical and sporadic group known -- all the way up to the Fischer-Griess Monster Group with $8\times10^{53}$ states (when represented as a transformation group)  -- but only two generators, so that such a group can live in a very simple environment indeed!

As we saw in the previous section, the probabilistic system contains a non-probabilistic submachine with many groups far more complicated than those instantiated by the corresponding noiseless mechanism. In the examples we have seen, however, their appearance is quickly overwhelmed by the uninteresting resets. Noise in the mechanism leads finally to a decay in the complexity of the emergent process. As mechanisms are combined, the noise in their resulting function breaks up their structure. 

\begin{figure}
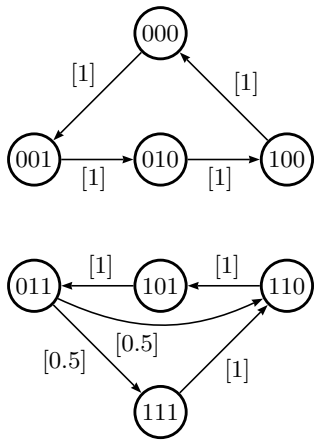

\VCDraw[0.7]{
\begin{VCPicture}{(-1,-3)(9,16)} 
\LargeState\State[000]{(4,12)}{000}

\State[001]{(0,8)}{001}
\State[010]{(4,8)}{010}
\State[100]{(8,8)}{100}

\State[011]{(0,4)}{011}
\State[101]{(4,4)}{101}
\State[110]{(8,4)}{110}

\State[111]{(4,0)}{111}

\EdgeR{000}{001}{[1]}
\EdgeR{001}{010}{[1]}
\EdgeR{010}{100}{[1]}
\EdgeR{100}{000}{[1]}

\EdgeR{101}{011}{[1]}
\EdgeR{110}{101}{[1]}
\VArcR{arcangle=-25}{011}{110}{[0.5]}

\EdgeR{011}{111}{[0.5]}
\EdgeR{111}{110}{[1]}

\end{VCPicture}
} 
\caption{A rare $\epsilon$-independent noisy automaton, on three bits, for which uncorrelated failures do not lead to convergence to a collection of resets. Such machines require at least partially noise-free behavior at the mechanism level.}
\label{rare}
\end{figure}

If the noise has the form of Eq.~\ref{prob}, can a system put zero weight on dissipative structures that combine into resets? Even a probabilistic choice on transformations of two ``counter-rotating'' cyclic groups will lead to a reset in $\pi(m)$ by Eq.~\ref{prob}, and the presence of an overlapping elementary collapsing and permutation are sufficient to produce the necessary irreversibility. Examples can be constructed where the resets apply only to part of the system; Fig.~\ref{rare} is one. Note, however, that the probabilistic behavior is confined to a subsystem -- and that this subsystem converges to a reset.

This is not, however, the final word, as we shall see in the following section, and the nature of the underlying mechanism can allow a system to avoid the asymptotic fate shown in Fig.~\ref{decay}.

\section{Dissipation, Irreversibility and the Varieties of Unreliability}
\label{correlated_noise}

A different way to alter the dissipative nature of noisy computation is to depart from Eq.~\ref{prob}. If the noisy mechanism that underlies a finite state machine is interpreted as a Markov process, Eq.~\ref{prob} is a general, but not the only, way to separate out the probabilistic behavior from the non-probabilistic structure. It contains implicit assumptions about the nature of the mechanism, and examination of alternative separations gives insight into the relationship between Markov processes and computation.

For the noisy-{\tt XOR} machine, with truth table $[\epsilon, (1-\epsilon), (1-\epsilon), \epsilon]$, Eq.~\ref{prob} produces a system with sixteen input letters. One could, however, imagine a decomposition, entirely equivalent if the state diagram of the machine is read instead as a Markov process, composed of only two letters, $a$ and $b$. The letter $a$ is drawn with probability $1-\epsilon$, and induces the transformation of Fig.~\ref{monoid}b; the letter $b$ is drawn with probability $\epsilon$ and induces the complementary transformation, \emph{i.e.}, that given by the truth table $[1,0,0,1]$. (Other decompositions, with intermediate alphabet sizes, can also be constructed, but we consider these two for simplicity.)

Any particular history of the machine's state transitions will be consistent with either decomposition. The differences emerge when one considers multiple histories at once: either by putting multiple pebbles on the machine, and moving them according to the transitions induced by the letters, or by putting the non-probabilistic sub-machine in different initial states and starting the Bernoulli process with the same ``seed.''

When considering multiple histories at once, it is seen that alternatives to the decomposition of Eq.~\ref{prob} amount to theories of correlated failures in the underlying mechanism. In the two-letter process, for example, whatever glitch occurs affects the output associated with each input in the same way -- by inversion -- whereas the 16-letter process allows for the possibility of a glitch that flips only some of the input-output relations of a process.

\begin{figure}
\includegraphics[width=3.25in]{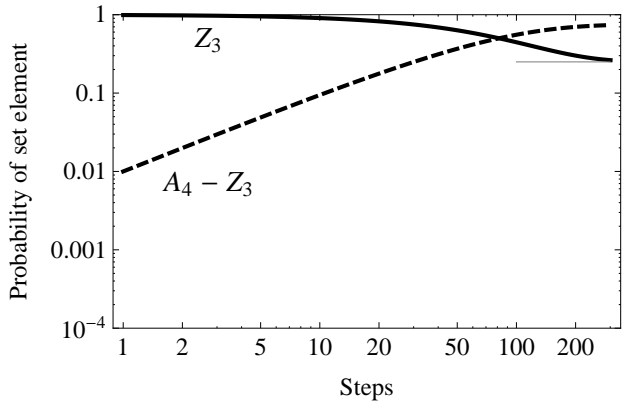}
\caption{The rise, and fall, of group structures in noisy Boolean systems with correlated unreliability as described in Sec.~\ref{correlated_noise}. In contrast to the standard, uncorrelated model of failure implied by Eq.~\ref{prob} and shown in Fig.~\ref{decay}, no irreversible transformations appear at all. However, knowledge of which reversible transformation the system has undergone becomes increasingly uncertain, and the distribution of possible transformations at long times becomes a uniform distribution over $A_4$, a subgroup of the full symmetric group, with elements of $Z_3$ appearing at the expected frequency of $1/4$ (short grey line.)}
\label{correlated_figure}
\end{figure}
The uncorrelated case of Eq.~\ref{prob}, known as the $\epsilon$-independent failure model, is that considered by Ref.~\cite{vN56}. In many cases -- such as shot noise in a biomolecular process, thermal noise in a semiconductor, or uncorrelated hidden variables in a group decision-making process -- such an assumption seems warranted. However, in other situations, noise properties are correlated, meaning that, for example, a failure that switches $f(x,y)$ to $\overline{f(x,y)}$ will necessarily switch $f(x^\prime,y^\prime)$ to $\overline{f(x^\prime,y^\prime)}$ as well.

In general, sources of these restricted processes include failures at ``high levels'' of the mechanism: the code that implements ${\tt XOR}$ is randomly flipped to code that implements ${\tt NOT}$-${\tt XOR}$. Another possibility is that of shared hidden variables, such that mechanisms that fail in a particular fashion in one state necessarily fail in the same way in any one of a particular subclass of such states. An example of such correlated noise in biological information processing, which could potentially lead to machines violating the implicit assumptions of Eq.~\ref{prob}, is found in mammalian motor and cortical neurons, where redundant signals from upstream processing arrive at different downstream dendrites~\cite{Jia:2010p19487,Priebe:2010p19415}. Fluctuations in membrane potentials are the primary source of neural variability~\cite{Calvin:1967p19490,Calvin:1968p19491}; the neural architecture is such that this synaptic noise is then shared among many downstream neurons leading to strong correlations in their spontaneous fluctuations~\cite{Lampl:1999p19489}.

Work in the theory of computation has, in parallel to these experimental discoveries, shown how the failure model implied by Eq.~\ref{prob} is insufficient to address questions of reliability when the possibility of randomized algorithms are taken into account. Ref.~\cite{Pippenger:1989p18957} describes alternatives to the $\epsilon$-independent model, but additional work, and richer theories of mechanism, are needed to accommodate the kinds of failures that violate Eq.~\ref{prob} in the ways described here.

The implications of correlated mechanism noise can be seen in Fig.~\ref{correlated_figure}, which shows the emergence of new group structures in the two-letter case described at the beginning of this section. Plotted on the same scale as Fig.~\ref{decay}, it shows how a restricted class of failures can avoid the transition to reset-dominance altogether. In particular, the existence of only one failure mode, where the circuit switches from ${\tt XOR}$ to ${\tt NOT}$-${\tt XOR}$, leads to a random walk whose asymptotic limit is the uniform distribution over the alternating group $A_4$ (even permutations on four elements), a subgroup of the full set of permutations.

The two-letter machine of Fig.~\ref{correlated_figure} provides an excellent example of the distinction between irreversibility and dissipation. No irreversible transformations are ever performed on the system states; the entropy remains constant and the system remains in equilibrium. Two pebbles placed on different initial states will never meet. However, the possibility of error -- which converges to a certainty -- means that a number of different histories are compatible with the present state, even when knowledge of the initial state is certain. The history of the organism can be decoded only with information about the internal states of its Bernoulli process.

\section{Conclusions}

Effective theories have proven to be a powerful tool for the investigation of the physical world; when combined with group theoretic arguments, they have allowed for principled exploration of phenomena where the underlying mechanisms remain unclear or unknown~\cite{Huggett:1995p19348}. We have shown here how effective theories can extend to irreversible, dissipative and out-of-equilibrium systems with a demonstration that relates an underlying mechanism (the Boolean circuits) to a higher-level decomposition that permits coarse-graining.

Irreversibility can occur ``by design,'' when certain environmental inputs are taken to be resets by the underlying mechanism. In natural systems, such resets may be useful in cases, for example, where radical changes in environment invalidate the count, or higher-level group property, the machine has arrived at.

But the related phenomenon of dissipation can also occur due to underlying noise in the mechanism. This fundamentally limits the lifetime of any particular group structure -- counts and more complicated memories can only be maintained for so long. Even when mechanisms are formally ``universal'' (for example, even when machines are assembled from gates in a universal basis), noise limits these universal properties, and leads to an association between (classes of) underlying mechanisms and emergent, effective theories.

Symmetry groups have played a central role in the physical sciences -- particularly in the study of effective field theories of elementary particles~\cite{Weinberg:1979p19325,joe,Georgi:1993p18122}. The presence of both irreversibility and dissipation in  biological, social and even computational~\cite{thermo} systems could potentially be taken as a sign that these methods will be ineffective in these new realms.

We find, instead, that such groups still have a role to play. When systems have irreversible mechanisms, hidden symmetries can still be found. Even when noise, and thus dissipation, is present, these symmetries may  survive, possibly even in enriched form, on short timescales. The planets are confined, by symmetry, to a single orbital plane for exceedingly long periods; despite the dissipation present, a bird flock may also show patterns, and group actions on those patterns, for short times before they fluctuate away.

\section{Acknowledgements}

I thank Chuck Stevens, Jay Garlapati, Attila Egri-Nagy and the attendees of the Santa Fe Institute Complex Systems Summer School 2011 for helpful conversations, and my two anonymous referees for their detailed comments and questions. I acknowledge the support of an Omidyar Postdoctoral Fellowship and National Science Foundation Grant EF-1137929, ``The Small Number Limit of Biological Information Processing.''

\section{Appendix}

Here we show the isomorphism between the left-shift powerset (LSP) semigroup (on $n$ bits) and the full transformation semigroup (FTS) (on $2^n$ elements.) It is not immediately obvious that the two semigroups are isomorphic; the LSP has (up to) $2^{(2^n)}$ generators (one generator for each truth table) and the FTS has $(2^n)^{(2^n)}$ elements. The FTS has three generators: the cyclic permutation, a swap of neighboring elements, and a elementary collapsing (all elements unchanged, except one element, which is mapped to its neighbor.) Only one of these (the cyclic permutation) appears in the generators of the LSP.
\begin{figure}
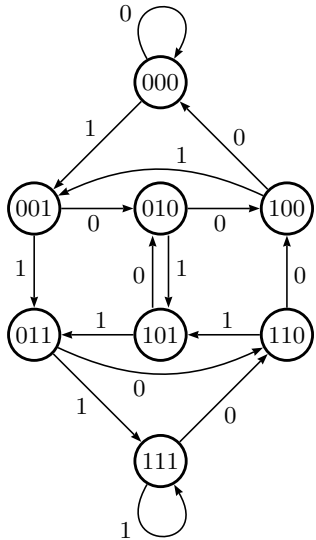

\VCDraw[0.7]{
\begin{VCPicture}{(-1,-3)(9,16)} 
\LargeState\State[000]{(4,12)}{000}

\State[001]{(0,8)}{001}
\State[010]{(4,8)}{010}
\State[100]{(8,8)}{100}

\State[011]{(0,4)}{011}
\State[101]{(4,4)}{101}
\State[110]{(8,4)}{110}

\State[111]{(4,0)}{111}

\EdgeR{000}{001}{1}
\LoopN{000}{0}
\EdgeR{001}{010}{0}
\EdgeR{010}{100}{0}
\VArcR{arcangle=-25}{100}{001}{1}
\ForthBackOffset \EdgeL{010}{101}{1} \EdgeL{101}{010}{0} \RstEdgeOffset
\EdgeR{001}{011}{1}
\EdgeR{101}{011}{1}
\EdgeR{110}{101}{1}
\EdgeR{110}{100}{0}
\VArcR{arcangle=-25}{011}{110}{0}

\EdgeR{011}{111}{1}
\EdgeR{111}{110}{0}
\EdgeR{100}{000}{0}
\LoopS{111}{1}

\end{VCPicture}
} 
\caption{The three-bit left-shift automaton.}
\label{3lsp}
\end{figure}

We shall show the isomorphism by showing that the three generators of the FTS can be found in the elements of the LSP. Since the FTS is the maximal semigroup possible, inclusion implies isomorphism. The reader may want to refer to Fig.~\ref{3lsp}; generators of the LSP can be constructed by choosing one of the two outgoing arrows for each node.

The cyclic permutation is easy to find in Fig.~\ref{3lsp} -- the path for the three-bit case runs $\{000,001,010,101,011,111,110,100,000\}$. A machine associated with this pattern for arbitrary bit length is the Linear Feedback Shift Register (LFSR; see Ref.~\cite{schneier96}.) The maximal LFSR relies solely on ${\tt XOR}$ to produce cycles over $2^n-1$ states (skipping the all-zero state), but any particular case can be augmented (using non-standard gates) to cycle over the full $2^n$ states.

Call the $2^n$ cycle $a$, and the $2^n-1$ cycle $b$. The two cycles can then be composed to produce a pair-swap -- in particular, $b^{(2^n-2)}a$ compose to make a pairwise swap of the first two elements. Since the LSP includes the cyclic permutation and the pair-swap, it includes the symmetric group.

It remains to show that the LSP includes an elementary collapsing. The cyclic permutation $a$ must include the transition $00\ldots0\rightarrow00\dots1$; if we change this to $00\ldots0\rightarrow00\dots0$, yielding another element of the LSP, then we have (after multiplying by $a^{(2^n-1)}$) an elementary collapse, and the proof of correspondence is complete.

%

\end{document}